\documentclass[superscriptaddress,aps,prl,showkeys,showpacs,twocolumn,10pt]{revtex4}
\usepackage[OT1,T1]{fontenc}
\usepackage{graphicx}
\usepackage{rotating}
\begin{document}

\title{On the Possibility of Dibaryon Formation near the $N^*(1440)N$
  threshold --- 
  the Isoscalar Single-Pion Production Revisited}
\date{\today}

\newcommand*{\PITue}{Physikalisches Institut, Eberhard--Karls--Universit\"at 
 T\"ubingen, Auf der Morgenstelle~14, 72076 T\"ubingen, Germany}
\newcommand*{\Kepler}{Kepler Center for Astro and Particle Physics, University
  of T\"ubingen, Auf der Morgenstelle~14, 72076 T\"ubingen, Germany}
\newcommand*{\INR}{Institute for Nuclear Research of the Russian Academy of
  Sciences, Moscow, Russia}

\author{H.~Clement}     \affiliation{\PITue}\affiliation{\Kepler}
\author{T.~Skorodko}    \altaffiliation[present address:
]{\Moscow}\affiliation{\PITue}\affiliation{\Kepler}
\author{E.~Doroshkevich} \affiliation{\INR}

\newcommand*{\Moscow}{Now at Moscow}

\begin{abstract}
The isoscalar single-pion production exhibits a broad bump in the
energy dependence of the total cross section, which does not correspond to the
usual opening of the $N^*(1440)$ production channel with subsequent pion
decay. In arxiv:2102.05575 it was interpreted as a narrow Breit-Wigner
structure, which leads in a sequential single-pion production process to a
possible explanation of the $d^*(2380)$ resonance. We demonstrate that such an
attempt fails already, when confronted with the data base for isoscalar
single-pion production. We investigate whether the observed bump structure
rather points to the 
formation of dibaryon states with $I(J^P) = 0(1^+)$ and $0(1^-)$ near the
$N^*(1440)N$ threshold. This situation would be similar to the situation at
the $\Delta(1232)N$ threshold, where the signature of a number of dibaryonic
resonances has been found. 
\end{abstract}


\maketitle

\section{Introduction}

In recent years many so-called exotic states have been observed in the charmed
and beauty quark sectors, both in mesons and baryons. These X, Y, Z and
pentaquark states appear as narrow resonances near particle thresholds
constituting weakly bound systems of presumably molecular character
\cite{CHreview}. In the following we
discuss the corresponding situation in the unflavored dibaryon sector,
which can be investigated by both  elastic nucleon-nucleon ($NN$) scattering
and $NN$-induced pion-production. Different from the flavored sector such
dibaryonic states decay into products, which usually contain unflavored
excitations of the nucleon. Since those have already a large intrinsic
hadronic width, such dibaryon excitations cannot be expected to be as narrow as
resonances in the flavored sector, even not near thresholds, where the phase
space for decay products is small.

After the recent observation of the - for a hadronic excitation - surprisingly
narrow dibaryon resonance $d^*(2380)$ with $I(J^P) = 0(3^+)$ in $NN$ scattering
\cite{np,npfull} and $NN$-induced two-pion production
\cite{MB,isofus,prl2011,pp0-,pn00,pn+-}, new measurements and re-investigations
revealed  
or reconfirmed evidences for various dibaryonic states near the $\Delta N$
threshold. The most pronounced resonance structure there is 
the one with the quantum numbers $I(J^P) = 1(2^+)$, mass $m \approx$ 2148 MeV
and width $\Gamma 
\approx$ 120 MeV, which is compatible with the width of $\Delta(1232)$. Its
structure in the $pp \leftrightarrow d\pi^+$ cross section coupled to the
$^1D_2$ $NN$-partial wave is known already since the fifties. Because
its mass is close to the nominal $\Delta N$ threshold of 2.17 GeV and its width
is compatible with
that of the $\Delta$ itself, its nature has been heavily debated in the
past, see, {\it e.g.},
Refs.~\cite{Shipit1,Shipit2,Ryskin,Kravtsov,Strakovsky,Arndt1,Arndt2,Arndt3,Hos1,Hos2}.
Its resonance behavior has been clearly observed separately in $\pi d$
\cite{Arndt3} and $pp$ \cite{Arndt2} scattering as well as in $pp \leftrightarrow
d\pi^+$ reaction \cite{Arndt1}. Also in the combined analysis of $pp$, $\pi d$
scattering and $pp \leftrightarrow d\pi^+$ reaction \cite{SAID} the resonance
effect in the $^1D_2$ $pp$-partial wave is apparent. For a recent review about
this issue see, {\it e.g.}, Refs. \cite{hcl,cpc}.

Recently also evidence for a resonance with mirrored quantum numbers, {\it
  i.e.} $I(J^P) = 2(1^+)$, mass $m =$ 2140(10) MeV and
width $\Gamma =$ 110(10) MeV has been published \cite{D21,D21long}. Due to its
isospin, this resonance can not couple directly to the $NN$ channel. However,
it can be produced associatedly in $NN$-induced two-pion production. It is 
remarkable that both these states as well as $d^*(2380)$
have been predicted already in 1964 by Dyson and Xuong \cite{Dyson} based on
$SU(6)$ multiplet considerations. More lately these states were calculated
also in a Faddeev treatment by Gal and Garcilazo \cite{GG,Gal} providing
agreement with experimental findings. 
These two states with mirrored quantum numbers, $I(J^P) = 1(2^+)$ and
$2(1^+)$, represent weakly bound states 
relative to the nominal $\Delta N$ threshold and hence are of presumably
molecular character with $N$ and $\Delta$ in relative $S$-wave --- a picture
supported by the Faddeev calculations of Refs. \cite{GG,Gal}.

Recently 
evidence has been presented for two further states, where the two baryons
$\Delta$ and $N$ are in relative $P$-wave: a state with 
$I(J^P) = 1(0^-)$, $m =$ 2201(5) MeV and $\Gamma =$ 91(12) MeV coupled to the
$^3P_0$ $NN$-partial wave as well as a
state with $I(J^P) = 1(2^-)$, $m =$ 2197(8) MeV and $\Gamma =$ 130(21) MeV
coupled to the $^3P_2$ $NN$-partial wave \cite{ANKE}. Whereas the  values for
the latter state agree with those obtained before already in SAID partial-wave
analyses \cite{SAID}, the $I(J^P) = 1(0^-)$ state was not known before, since
it is forbidden in the well-investigated two-body
reaction $pp \rightleftharpoons d\pi^+$. The masses of these $P$-wave
resonances are slightly above the nominal $\Delta N$ threshold, which is
understood as being due to the additional orbital motion \cite{ANKE}.

There is
evidence for the existence of still further states like another $\Delta N$
$P$-wave state with $I(J^P) = 1(3^-)$, $m =$ 2183 MeV and $\Gamma =$ 158 MeV
coupled to the $^3F_3$ $NN$-partial wave \cite{SAID}. However, the
experimental situation there is not yet as clear \cite{hcl}.

Platonova and Kukulin demonstrated recently that both cross section and
polarization observables of the $pp \to d\pi^+$ reaction \cite{Kukulindpi+} as
well as the participating dominant $NN$-partial waves \cite{KukulinNN} can be
described consistently on a quantitative level, if dibaryon
resonances in the $^3P_2$, $^1D_2$ and $^3F_3$ $NN$-partial waves are included.
As already pointed out in previous studies \cite{Igor}, it is concluded that
these partial waves contain both genuine resonant parts (dibaryon resonances) as
well as pseudoresonant parts (due to the $\Delta N$ intermediate state).

Recent photoproduction experiments carried out at ELPH, Tohoku, and ELSA, Bonn,
suggest that
also at thresholds of higher-lying baryon excitations dibaryonic structures
are formed \cite{Ishikawa,Jude}. According to their $\gamma d \to d\pi^0\pi^0$
measurements the observed structures in the so-called second and third
resonance region do not represent quasi-free processes for baryon excitations,
but rather constitute dibaryonic excitations at 2.47 and 2.63 GeV, respectively.

In the following we investigate, whether the scenario of dibaryonic resonances
near baryon excitation thresholds finds also some repetition near the
$N^*(1440)N$ threshold. In a preceeding work \cite{NstarN} it was demonstrated
that the $^1S_0$ and $^3S_1$ $NN$-partial waves can be well described, if
dibaryon resonances with $I(J^P) = 1(0^+)$ and $0(1^+)$ near the $N^*N$
threshold are postulated, for which also suggestive experimental evidence was
presented. The evidences for the $I(J^P) = 1(0^+)$ state will be reconsidered in
this work. 

There is yet another reason to look in more detail into the isoscalar
single-pion production. Recently an article \cite{EO} appeared claiming that
sequential single-pion production is able to explain the $d^*(2380)$ peak in
the $np \to d\pi^+\pi^-$ reaction by the particular two-step process $np(I=0) \to
(pp)\pi^- \to (d\pi^+) \pi^-$. Since the second process proceeds dominantly
via the incident $^1D_2$ $pp$-partial wave, it is assumed in that work
silently that the 
$pp-pair$ emitted in the first reaction step is dominantly in this
particular partial wave. In addition, the authors fit the energy dependence of
the observed isoscalar single-pion production cross section by a Breit-Wigner
resonance ansatz with a width as narrow as 70 MeV --- without giving any
explanation for such a strikingly narrow resonance structure --- which by
itself would be an exiting and unique structure never observed before in any
single-pion production. Their fit is also in
conflict with the results in Refs. \cite{NNpi,NNpicorr}, where the observed
energy dependence was fitted by a width of about 150 MeV. We take these claims
as yet another reason to reinspect thoroughly the experimental situation in
the isoscalar single-pion production.

\section{Experimental Situation in Single-Pion Production}
\subsection{The purely isovector reaction $pp \to pp\pi^0$}

The $\pi^0$-production in $pp$ collisions has been measured by many groups
with a number of different equipments
\cite{bugg,shim,eis,Bys,Sarantsev,Sarantsev1,AS,Rappenecker,hades,Flaminio}. Fig.~1 shows the
resulting total cross section from threshold up to $T_p$ = 1.5 GeV ($\sqrt s$
= 2.6 GeV). Since we are interested here mainly in the region, where the cross
section starts to saturate, we do not plot the energy dependence of the
total cross section in logarithmic scale as usually done, but in linear scale,
in order to focus on the situation of available data in the region of interest. 

Whereas the data in the near-threshold region exhibit a rather
consistent behavior of a strongly increasing cross section, the available
database beyond $T_p$ = 0.8 GeV ($\sqrt s$ = 2.2 GeV) displays quite
some scatter in the region, where the cross section starts to flatten
out. There are essentially two groups of measurements, which do not coincide
well within their uncertainties. The one group favors cross section values
around 4 mb, the other one favors values around 4.5 mb. The WASA-at-COSY data
\cite{NNpi} were normalized to the average of previous measurements in this
region, which is well represented by the result of Ref. \cite{shim} at $\sqrt
s$ = 2.35 GeV. The WASA-at-COSY data exhibit a flat energy dependence in the
region of interest. 

\begin{figure} [t]
\centering
\includegraphics[width=0.9\columnwidth]{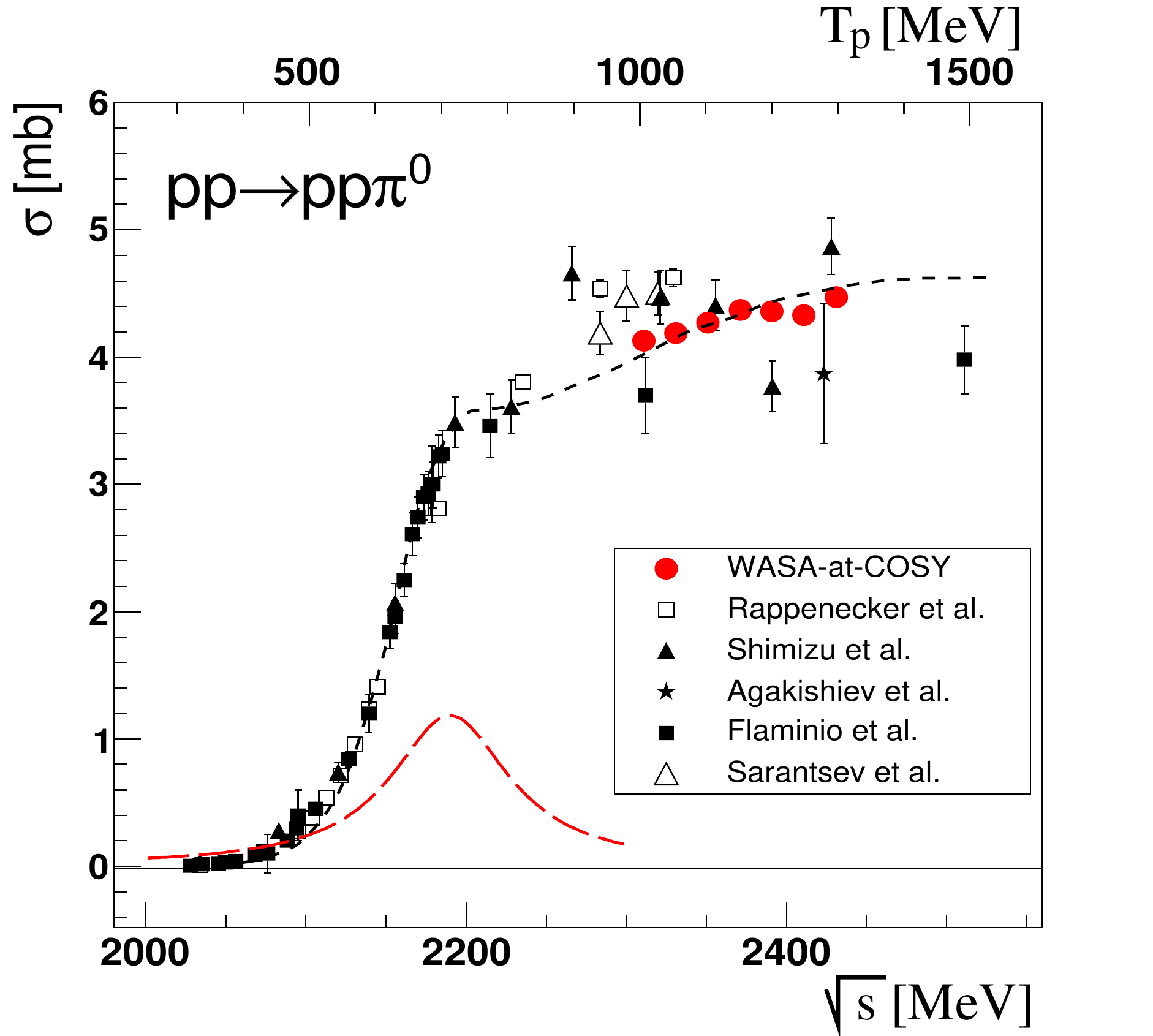}
\caption{\small (Color online) 
  Energy dependence of the total cross section for
  the $pp \to pp\pi^0$ reaction. Red filled circles denote the results
  from WASA-at-COSY \cite{NNpi}. Other symbols give
  results from earlier work \cite{shim,Bys,Sarantsev,Sarantsev1,AS,Rappenecker,hades,Flaminio}. The
  data points of Ref.~\cite{bugg} at $T_p = $ 970 MeV and Ref.~\cite{eis} at
  $T_p =$ 1480 MeV are included in the
  collection of Flaminio et al.~\cite{Flaminio}. The 
  short-dashed line 
  represents a calculation for $t$-channel $\Delta$ and $N^*(1440)$ excitation
  in the framework of the Valencia model \cite{Luis} - rescaled by a factor
  0.98. The long-dashed curve shows a Lorentzian fitted to the data in the $\Delta
  N$ region representing phenomenologically the contributions from the isovector
  $s$-channel dibaryon excitations $I(J^P) = 1(0^-), 1(2^-)$, $1(2^+)$ and
 $1(3^-)$ fed by the $^3P_0, ^3P_2$, $^1D_2$ and $^3F_3$ $NN$-partial
  waves. The dash-dotted curve gives the superposition of both contributions
  providing thus the full isovector cross section.
}
\label{fig1}
\end{figure}

The main physics in the region of interest may be inferred from Fig.~3 of
Ref. \cite{NNpi}, where differential cross sections accumulated
by the WASA-at-COSY experiment are shown over the energy region $T_p$ = 1.0 -
1.35 GeV ($\sqrt s$ = 2.3 - 2.45 GeV).
%Since in a three-body final state there four independent
%differential observables, we choose to show in Fig.~2 the differential
%distributions for the center-of-mass (c.m.) angles for protons and pions
%denoted by $\Theta_p^{c.m.}$ and  $\Theta_{\pi^0}^{c.m.}$, respectively, as
%well as for the invariant $p\pi^0$ and $pp$ masses, denoted by $M_{p\pi^0}$
%and $M_{pp}$, respectively. The results from
%$Ref.~\cite{iso} are shown by open symbols, the new ones from the reanalysis
%of this work by the full circles. The differences between both analyses are
%quite small. They may be taken as a measure for systematic uncertainties due to
%the different analysis methods.
All differential distributions deviate largely from pure phase-space
distributions. The $M_{p\pi^0}$ spectrum
exhibits a pronounced peak resulting form the excitation of the $\Delta(1232)$
resonance in the course of the reaction process. The strongly anisotropic
proton angular distribution is in accord with a peripheral reaction process
and the also anisotropic  pion angular distribution may be associated with the
$p$-wave decay of the $\Delta$ excitation.
%The solid lines give a model
%calculation of the $t$-channel $\Delta$ excitation and its decay in the course
%of the $pp \to pp\pi^0$ reaction process by utilizing the Valencia code for
%pion production \cite{Luis}. The dotted lines show the same calculations, but
%with inclusion of the $N^*(1440)$ excitation contributing 0.3 mb ??? to the
%total cross section. 

In an reanalysis of the WASA-at-COSY data \cite{NNpi} we confirm the
published differential cross sections within their quoted uncertainties. So
there no need to show them here again. Instead we show the Dalitz plot of the
$pp$-invariant mass-squared $M_{pp}^2$ 
versus the $p\pi^0$-invariant mass-squared $M_{p\pi^0}^2$ in Fig.~2.
The data from our reanalysis are shown on the top and a model calculation for
$\Delta$ excitation is displayed on the bottom. In both data and calculation
the vertical band for $\Delta$ 
excitation is clearly seen as well as its reflection due to the fact that we
have two identical protons, where the $\Delta$ excitation can happen in either
one. 

\begin{figure} %[t]
\begin{center}
\includegraphics[width=0.99\columnwidth]{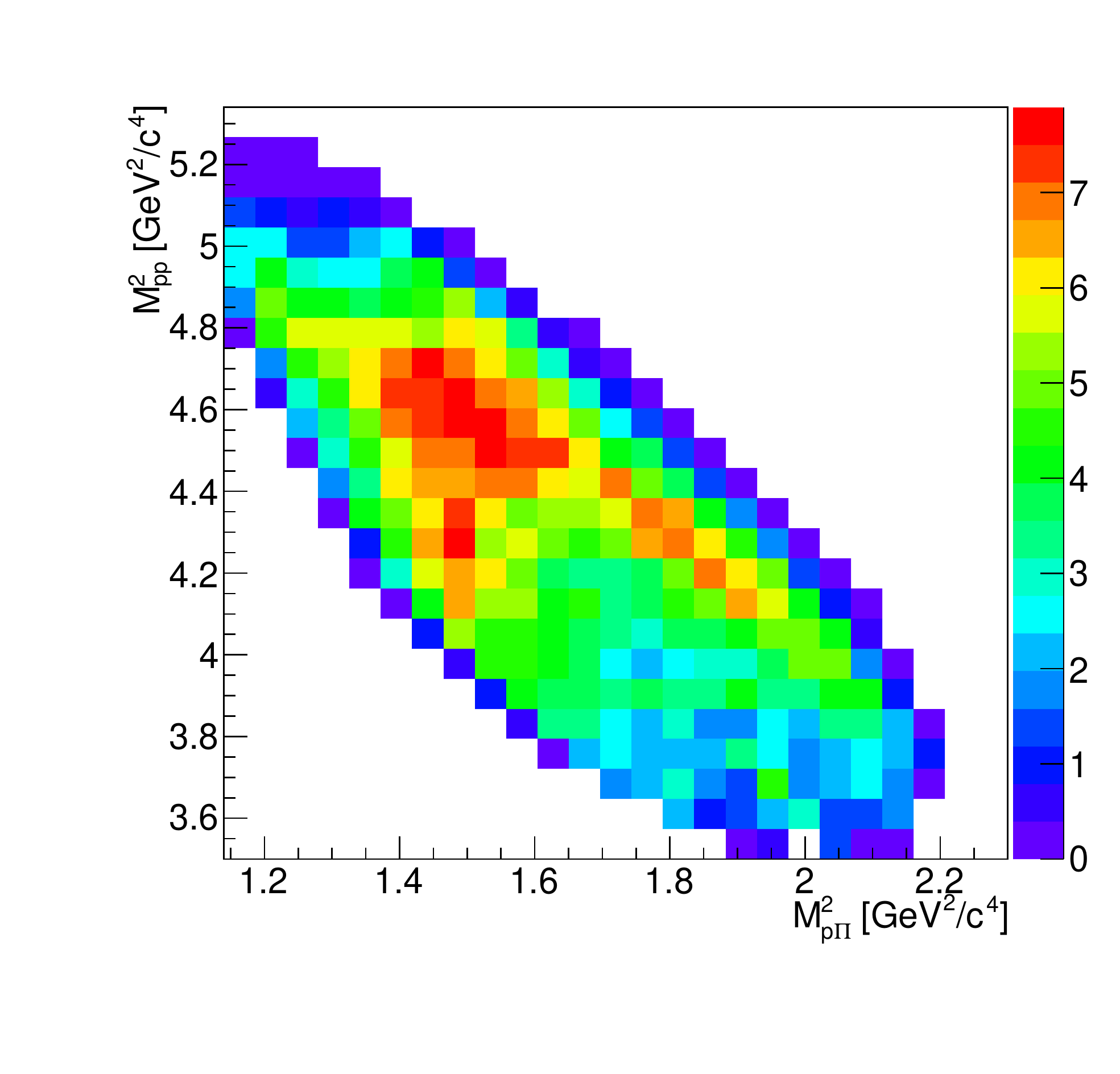}
\includegraphics[width=0.99\columnwidth]{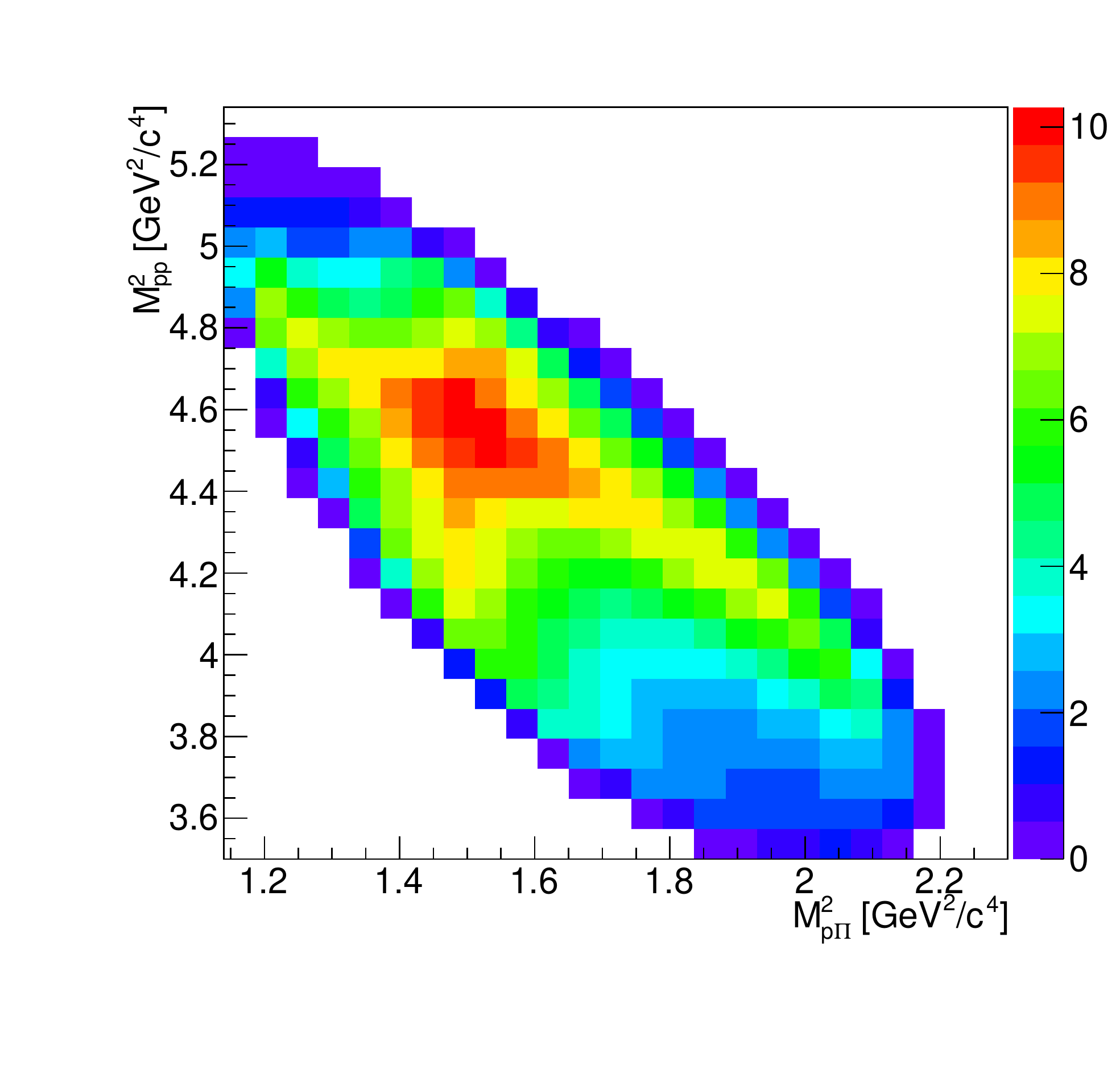}
\caption{(Color online) 
 Dalitz plot of the $pp$-invariant mass-squared $M_{pp}^2$ versus the
 $p\pi^0$-invariant mass-squared $M_{p\pi^0}^2$ for the energy bin
$\sqrt s$ = 2.40 - 2.42 GeV of the $pp \to pp\pi^0$ reaction. On the top the
data from our reanalysis are shown and on the bottom a model calculation for
$\Delta$ excitation is displayed. The intensity distribution is color coded in
the usual way in a linear scale with violet and red colors denoting the lowest
and the highest intensities, respectively. 
}
\label{fig2}
\end{center}
\end{figure}

%\begin{figure} %[t]
%\begin{center}
%\includegraphics[width=0.49\columnwidth]{invppi_pppi0.eps}
%\includegraphics[width=0.49\columnwidth]{invpp_pppi0.eps}
%\includegraphics[width=0.49\columnwidth]{cospi_pppi0.eps}
%\includegraphics[width=0.49\columnwidth]{cosp_pppi0.eps}
%\caption{(Color online) 
%  Differential distributions of the $pp \to pp\pi^0$ reaction at $T_p$ =
%  1.2. GeV for invariant-masses $M_{p\pi^0}$ (top left) and $M_{pp}$ (top right%)
%  of $p\pi^0$ and $pp$ subsystems, respectively, as well as  for the
%  c.m. angles of neutral pions $\Theta_{\pi^0}^{c.m.}$ (bottom left) and
% protons $\Theta_p^{c.m.}$ (bottom right). The hatched histograms indicate
%  systematic uncertainties due to the restricted phase-space coverage of the 
%  data. The light-shaded (yellow) areas represent pure 
%  phase-space distributions, the solid lines are calculations of
%  $\Delta(1232)$ and $N^*(1440)$ excitations by $t$-channel meson exchange --
%  normalized in area to the data.
%}
%\label{fig2}
%\end{center}
%\end{figure}

All data are very well described by assuming just $\Delta$ excitation in the
reaction process. Inclusion of a small Roper contribution does not change the
fit to the data noticeably. But the fit to the data starts to
deteriorate markedly, if the Roper contribution exceeds 0.4 mb in the total
cross section. This finding may serve us as an upper limit for the isovector Roper
contribution in the  $pp \to pp\pi^0$ reaction. We note that the observations
in the differential spectra from WASA-at-COSY are consistent with those
obtained in Refs. \cite{Sarantsev,AS} at lower energies.

\subsection{The isospin-mixed reaction $pn \to pp\pi^-$}

For this reaction there are much less measurements due to the need for an
effective neutron beam or target. Some experiments were conducted by 
utilizing the quasifree reaction process in the collision of deuterons with
protons by using either a deuteron beam hitting a hydrogen bubble
chamber\cite{Tsuboyama} or a proton beam hitting a deuteron bubble chamber
\cite{Dakhno,Brunt}. The measurements of Refs. \cite{Tsuboyama,Dakhno} are
over a wide energy range, their resulting cross sections are in good
agreement to each other in the overlap region.

Other experiments used a
dedicated neutron beam produced in a first scattering process by proton
collisions on a deuteron target, where the produced neutron beam was directed
either on a hydrogen bubble chamber \cite{AS} or on a liquid hydrogen target
\cite{Thomas,Kleinschmidt}. In the latter the isospin-mirrored reaction $np
\to nn\pi^+$ was measured in the near-threshold region.

The total cross sections obtained in the measurements are shown in Fig.~3 in
linear scale. There is good agreement between the WASA-at-COSY measurements
\cite{NNpi} and previous results from Refs.~\cite{Tsuboyama,Dakhno,Brunt,Bys}
with the exception of the data point at $T_P = $ 1.17 GeV ($\sqrt s =$ 2390
MeV) from Ref.~\cite{Tsuboyama}, which is far off from the other experimental
  results. 

\begin{figure} 
\centering
\includegraphics[width=0.99\columnwidth]{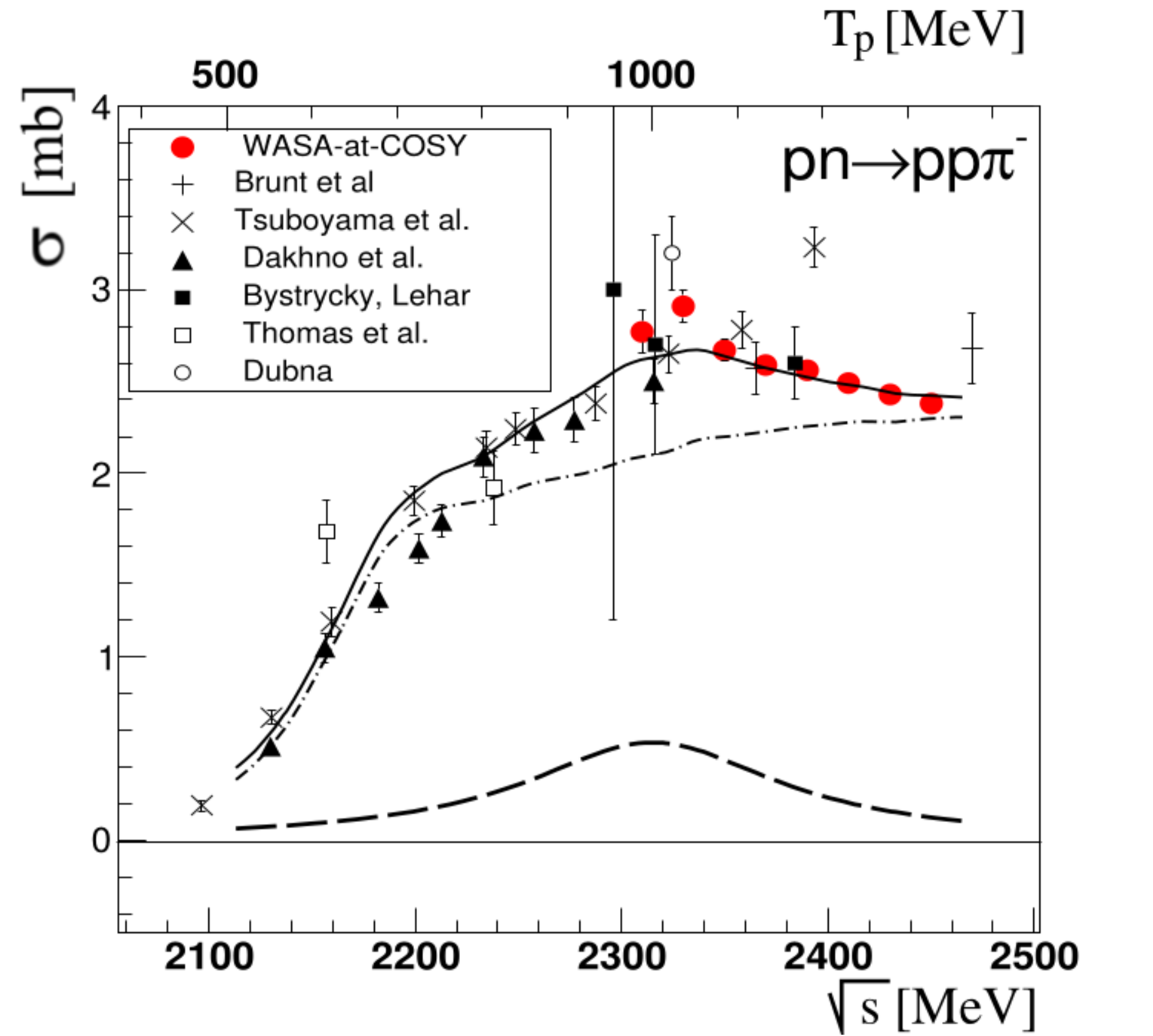}
\caption{\small (Color online) 
Energy dependence of the total cross section in dependence for
  the $pn \to pp\pi^-$ reaction. Red filled circles denote the results
  from WASA-at-COSY \cite{NNpi}. Other symbols give
  results from earlier work \cite{Dakhno,Bys,Thomas,Dubna,Tsuboyama,Brunt}. The dash-dotted line
  gives the purely isovector contribution obtained by the dash-dotted line in
  Fig.~1 with the absolute scale being reduced by a factor of two. Adding
  the Lorentzian from Fig.~6 (dashed curve) - divided by a factor of three for
  the representation of the isoscalar contribution in this channel - results
  in the solid curve.  
}
\label{fig3}
\end{figure}

As demonstrated in Ref.~\cite{NNpi} the differential distributions of the $pn
\to pp\pi^-$ reaction can no longer be described just the $\Delta$
excitation, but necessitate also a substantial Roper excitation.
This is also borne out in the Dalitz plot displayed in Fig.~4 for the $pn \to
pp\pi^-$ reaction.

\begin{figure} %[t]
  \begin{center}
\includegraphics[width=0.99\columnwidth]{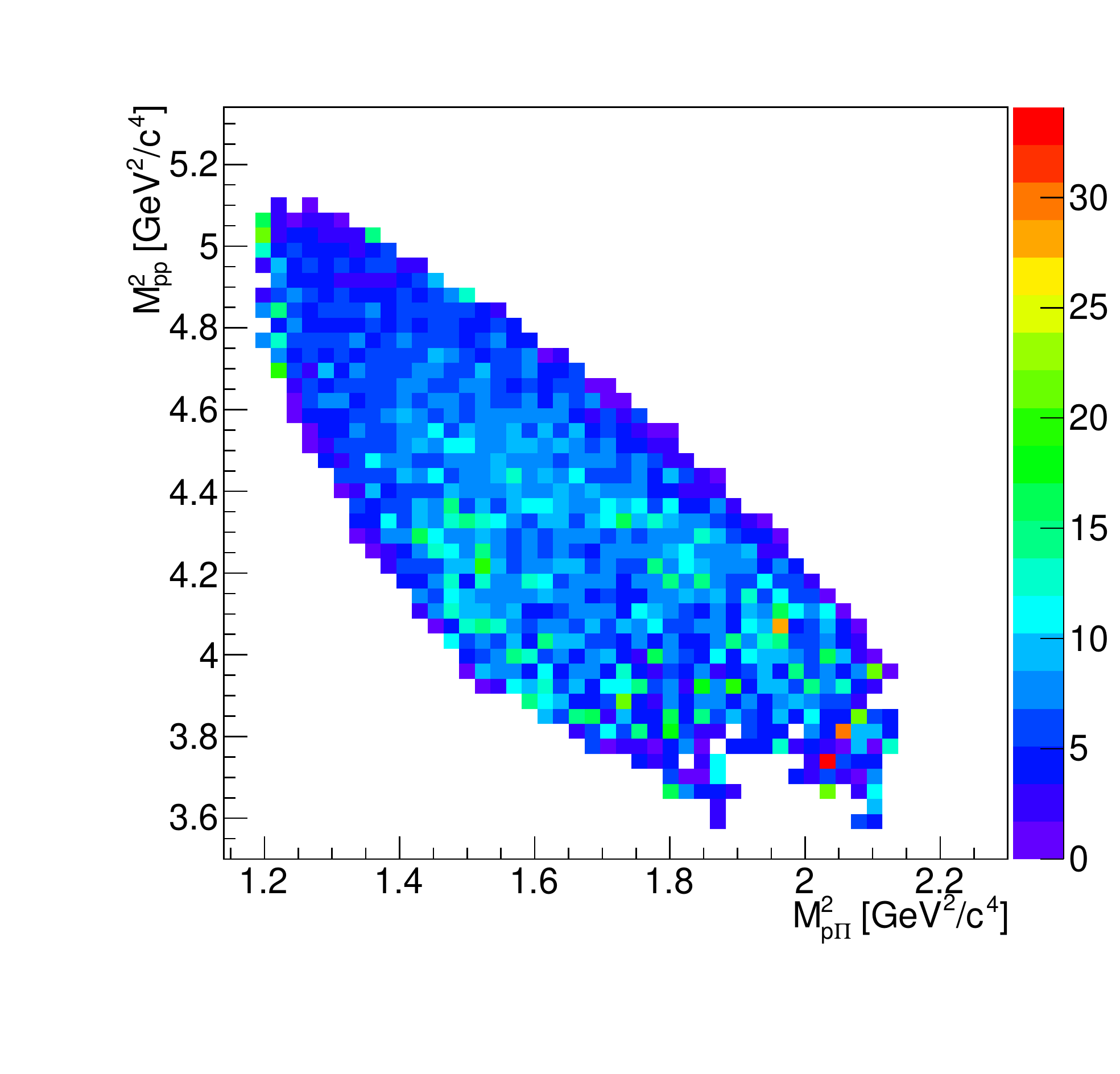}
\includegraphics[width=0.99\columnwidth]{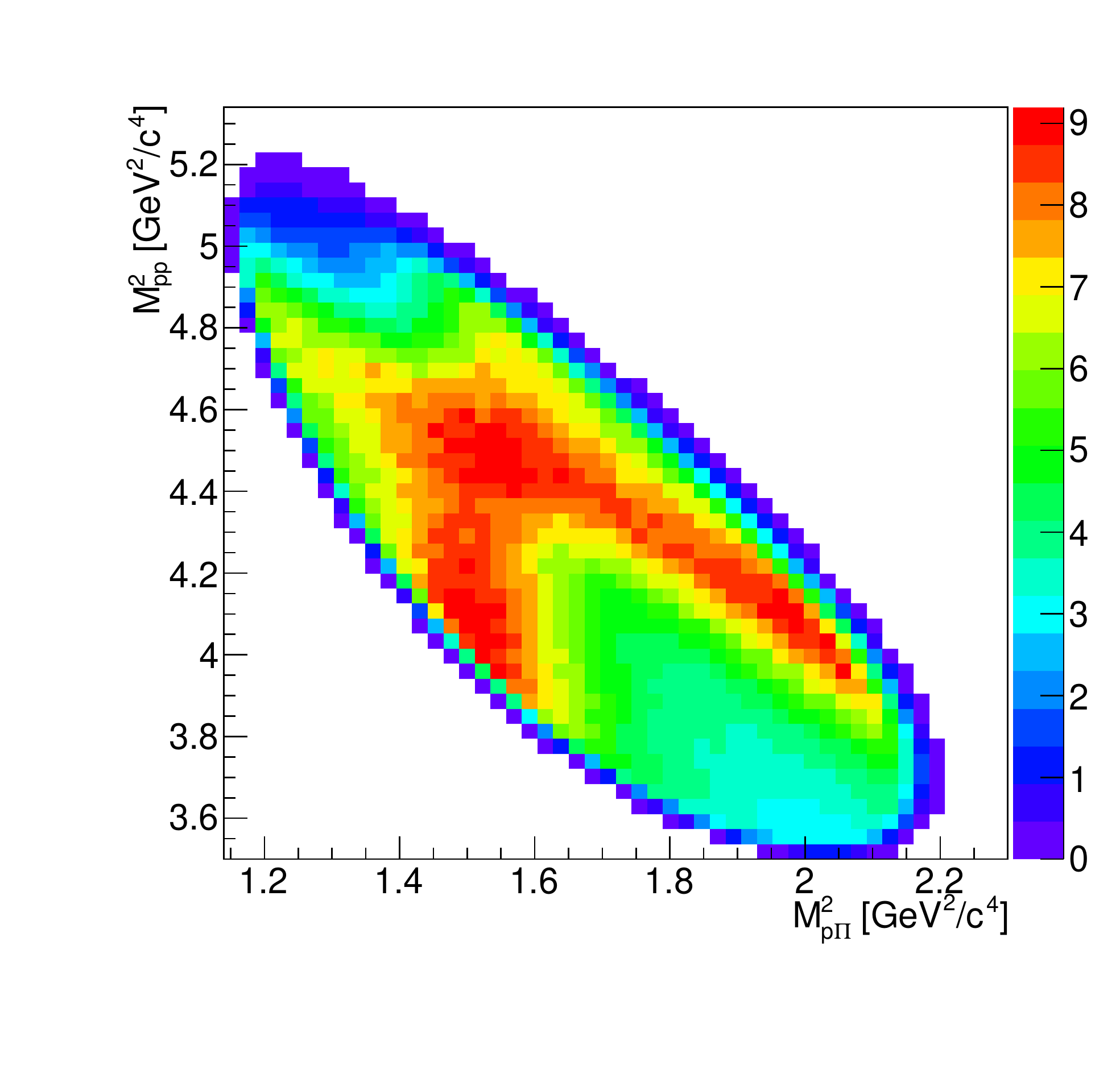}
\caption{(Color online) 
 Dalitz plot of the $pp$-invariant mass-squared $M_{pp}^2$ versus the
 $p\pi^-$-invariant mass-squared $M_{p\pi^-}^2$ for the energy bin
$\sqrt s$ = 2.40 - 2.42 GeV of the $pn \to pp\pi^-$ reaction. On the top the
data from our reanalysis are shown and on the bottom a model calculation for
$\Delta$ and Roper excitations is displayed. The intensity distribution is
color coded in the usual way in a linear scale with violet and red colors
denoting the lowest and the highest intensities, respectively.
}
\label{fig4}
\end{center}
\end{figure}

%\begin{figure} %[t]
%\begin{center}
%\includegraphics[width=0.49\columnwidth]{invppi_pppim.eps}
%\includegraphics[width=0.49\columnwidth]{invpp_pppim.eps}
%\includegraphics[width=0.49\columnwidth]{cospi_pppim.eps}
%\includegraphics[width=0.49\columnwidth]{cosp_pppim.eps}
%\caption{(Color online) 
%   The same as Fig. 3, but for the $pn \to pp\pi^-$ reaction.
%}
%\label{fig3}
%\end{center}
%\end{figure}

\subsection{The isoscalar single-pion production}

The isoscalar part of the $NN$-induced single-pion production cannot be measured
directly. It rather has to be deduced from a combination of various
single-pion production measurements. Most common is the comparison of the
total cross sections for the $pp \to pp\pi^0$ and $np \to pp\pi^-$ reaction
channels by assuming isospin invariance:

\begin{eqnarray}
  \sigma_{pn \to NN\pi}(I=0) = \frac{3}{2} (2 \sigma_{pn \to pp\pi^-} -
  \sigma_{pp \to pp\pi^0}),
\end{eqnarray}

where $\sigma_{pn \to NN\pi}(I=0)$ denotes the isoscalar $np$-induced
single-pion production cross section \cite{NNpicorr,Dakhno,Bys}. Results
obtained by use of this 
method are shown in Fig.~5 by solid dots \cite{NNpi}, solid triangles
\cite{Dakhno} solid squares \cite{Sarantsev} and open triangles
\cite{Rappenecker}.

Since we have the difference
of two nearly equally sized values in eq.~(1), the relative uncertainty in the
absolute normalization of the two cross section values leads to a large
uncertainty in the resulting isoscalar cross section. This explains also the
large scatter in the obtained results. Nevertheless, all data are consistent
with a increasing cross section from threshold up to $\sqrt s \approx$ 2300
MeV and leveling off there. The WASA-at-COSY data show that the cross section
starts falling at subsequent higher energies. 
%But there is also a systematic problem
%with the absolute values of the total cross section of the $pp\pi^0$
%channel, see also discussion below. The values used by Ref.~\cite{Dakhno} are
%in general about 10$\%$ 
%smaller than those used by  Ref. \cite{Sarantsev}. By using the same values
%for the total cross section of the $pp\pi^-$ channel this results via eq.~(1)
%to lower values for the isoscalar cross section of up to 50$\%$ . 

An alternative to this difference method given by Eq.~(1) has been employed in
Ref.~\cite{AS}. There all differential distributions obtained in
bubble-chamber measurements of the $pp \to pp\pi^0$ and $np \to pp\pi^-$
reactions have been subjected to a partial-wave analysis (PWA). The interference
between isovector and isoscalar amplitudes as it shows up in differential cross
sections, in particular in angular
distributions, provides a discrimination between isoscalar and isovector
contributions in the partial-wave analysis. Hence the results of this work
appear to be particularly reliable.
%Also, in contrast to the difference
%method given in eq.~(1) these results do not rely basically on isospin
%conservation.
They are given in Fig.~6 by the hatched band, where the bandwidth denotes the
uncertainty of that analysis. The band rises with rising energy reaching a
peak around $\sqrt s \approx 2.30$ GeV  and starts falling in height
thereafter. This latter feature agrees with the trend observed by the
WASA-at-COSY data \cite{NNpi}, only that the WASA-at-COSY values are 
higher by about 30$\%$ in the overlap region, which, however, is well within
their uncertainty in absolute scale \cite{NNpi}.

In the following we show that the large scatter in the experimental results for
$\sigma_{pn}(I=0)$ can be easily cured by a slight renormalization of the
various data samples well within their quoted uncertainties.
Inspection of Eq.~1 shows that an uncertainty $\delta\sigma$ in the absolute
magnitude of the $pp \to pp\pi^-$ cross section relative to that of the $pn
\to pp\pi^0$ cross section enters linearly in eq.~1 by a term $3\delta\sigma =
3\sigma_{pn \to pp\pi^-} (\delta\sigma / \sigma_{pn \to 
  pp\pi^-})$, which causes essentially a baseline shift in the deduced data
for $\sigma_{pn \to NN\pi}(I=0)$ in the region, where $pp \to pp\pi^0$ and $pn
\to pp\pi^-$ cross sections level off. So already a 1$\%$ change in the
normalization of $\sigma_{pn \to pp\pi^-}$ , {\it i.e.} $\delta\sigma /
\sigma_{pn \to pp\pi^-} = 1\%$,
% relative to $\sigma_{pn \to pp\pi^0}$
leads to a shift of $\sigma_{pn \to NN\pi}(I=0)$ by $3\delta\sigma
\approx$ 0.08 mb for $\sqrt s >$ 2.25 GeV. 
Hence, in order to achieve agreement
between PWA and WASA-at-COSY results it suffices to change the relative normalization
between  $pp \to pp\pi^0$ and $pn\to pp\pi^-$ cross sections of the WASA data
by 4$\%$ leading to a shift of about 0.3 mb. Such a renormalization of the WASA
results is well within the uncertainty of 7$\%$ in the relative
normalization between $pp \to pp\pi^0$ and $pn\to pp\pi^-$ cross sections
quoted in Ref.~\cite{NNpi}. Similarly, we may obtain reasonable overlap of the
PWA results with those of Refs. \cite{Sarantsev} and \cite{Rappenecker}, if we
renormalize those by 3$\%$ and 4$\%$, respectively, in their relative
normalization between $pp \to 
pp\pi^0$ and $pn\to pp\pi^-$ cross sections. Again, this is well within the
uncertainties there. In particular we see that by such a renormalization the
results of Ref.~\cite{Sarantsev} get in practical perfect overlap with the
uncertainty band of the PWA results \cite{AS}.

The renormalized data of Refs.~\cite{Sarantsev,Rappenecker} and
WASA-at-COSY~\cite{NNpi}   are compared with the PWA results
in Fig.~6, where they exhibit now a very consistent structure of an isoscalar
cross section rising from threshold up to about 2.3 GeV and declining
thereafter. This structure can be well described
by a Breit-Wigner shape having a width of 150(20) MeV and peaking at
2.31(1) GeV --- in accordance with the results reported in
Refs.~\cite{NNpi,NstarN}. Also the results from Refs.~\cite{Dakhno,Tsuboyama}
fit reasonably well, without any need for renormalization.
Only the highest energy point from Ref.~\cite{Tsuboyama} at $\sqrt s =$
2390 MeV is far away from the trend of the other data. The reason for this
lies in the much too large cross section $\sigma_{pn \to pp\pi^-}$ obtained by
Ref.~\cite{Tsuboyama} at that energy --- see Fig.~3.

%We note that in Ref.~\cite{EO} an isospin violation of 5$\%$ was assumed in
%each of the two terms in eq.~(1) leading to a systematic change of the
%absolute values for the isoscalar cross section again by about
%30$\%$. Possibly such an effect could explain, too, the discrepancy in scale
%between the WASA results and the results of Ref.~\cite{AS}.

As already noted in the introduction, Ref.~\cite{EO} has fitted the energy
dependence of the isoscalar cross section by a narrow Breit-Wigner form, which
by itself would point to a very spectacular resonance phenomenon in this
channel. But based on all the experimental data discussed here 
there is no way to obtain a Breit-Wigner fit with a width as
narrow as 70 MeV and peaking at 2.33 GeV. In Ref. \cite{EO} this could be
achieved only by enlarging the uncertainties of the WASA-at-COSY results
enormously by adding in quadrature a large systematic error due to isospin
violation. Such a procedure is by no means justified, since the
isospin violation is not fluctuating randomly from energy point to energy point
and hence does not behave like statistical uncertainties. Therefore it
cannot be added to them. Isospin violation rather affects just the
absolute scale of the isoscalar cross section shifting the data only in common
up or down in scale --- in the same way as discussed above with regard to the
relative normalization between the cross sections for the $pp \to pp\pi^0$ and
$np \to pp\pi^-$ reactions.

\begin{figure} 
\centering
\includegraphics[width=0.99\columnwidth]{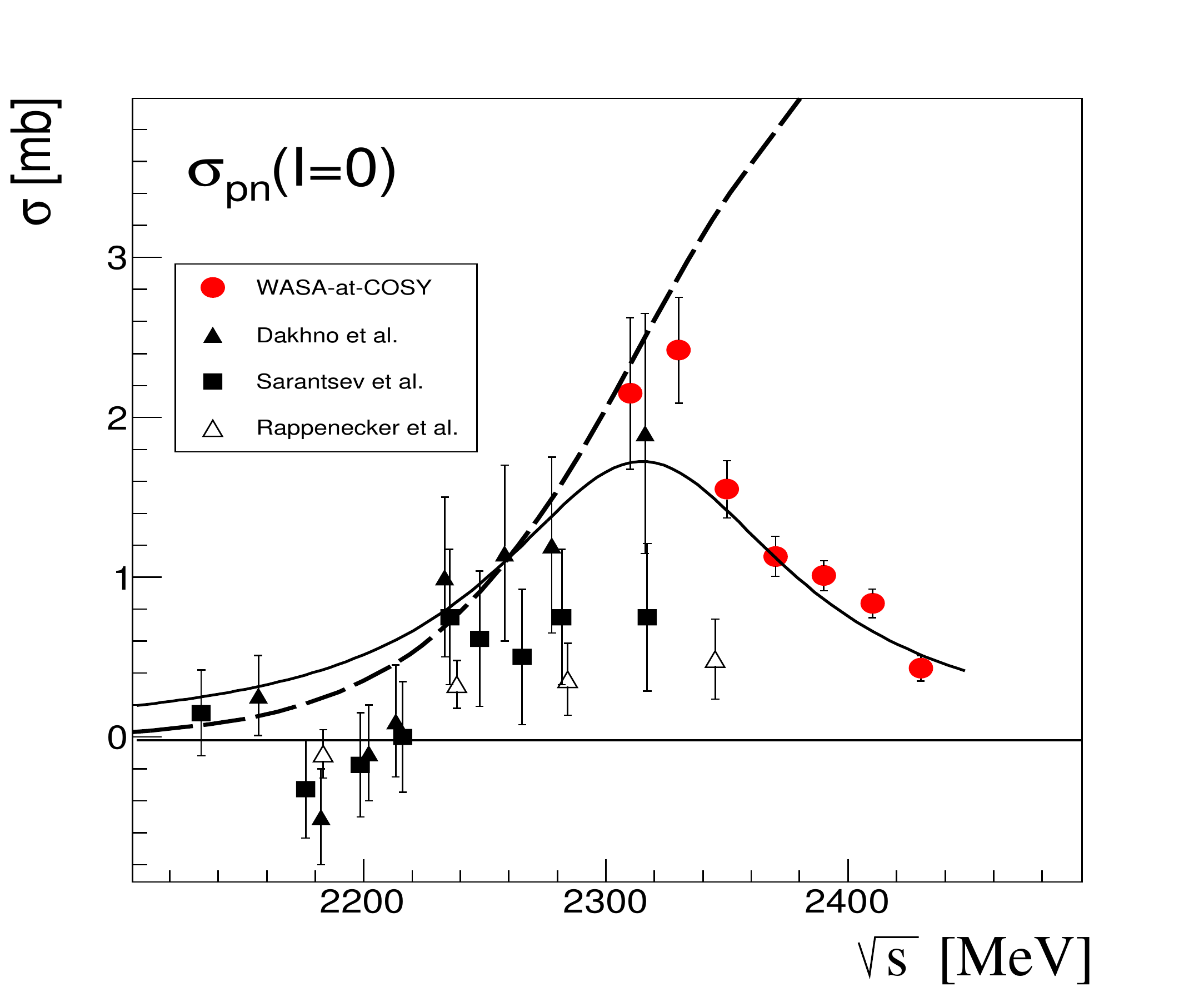}
\caption{\small (Color online) 
The $pn$-induced isoscalar single-pion production cross section based on eq. (1) in dependence
of the total c.m. energy $\sqrt s$. Shown are the recent results from
WASA-at-COSY \cite{NNpi,NNpicorr} (solid circles) together with earlier results
from Ref.~\cite{Dakhno} (solid triangles), Ref.~\cite{Sarantsev} (solid
squares) and Ref.~\cite{Rappenecker} (open triangles). The dashed line shows the
expected energy dependence based on $t$-channel Roper excitation \cite{Luis}
adjusted in
height arbitrarily to the data point at $\sqrt s$ = 2260 MeV. The solid lines
represents a Lorentzian with $m$ = 2315 MeV and $\Gamma$ = 150 MeV. 
}
\label{fig4}
\end{figure}

\begin{figure} 
\centering
\includegraphics[width=0.99\columnwidth]{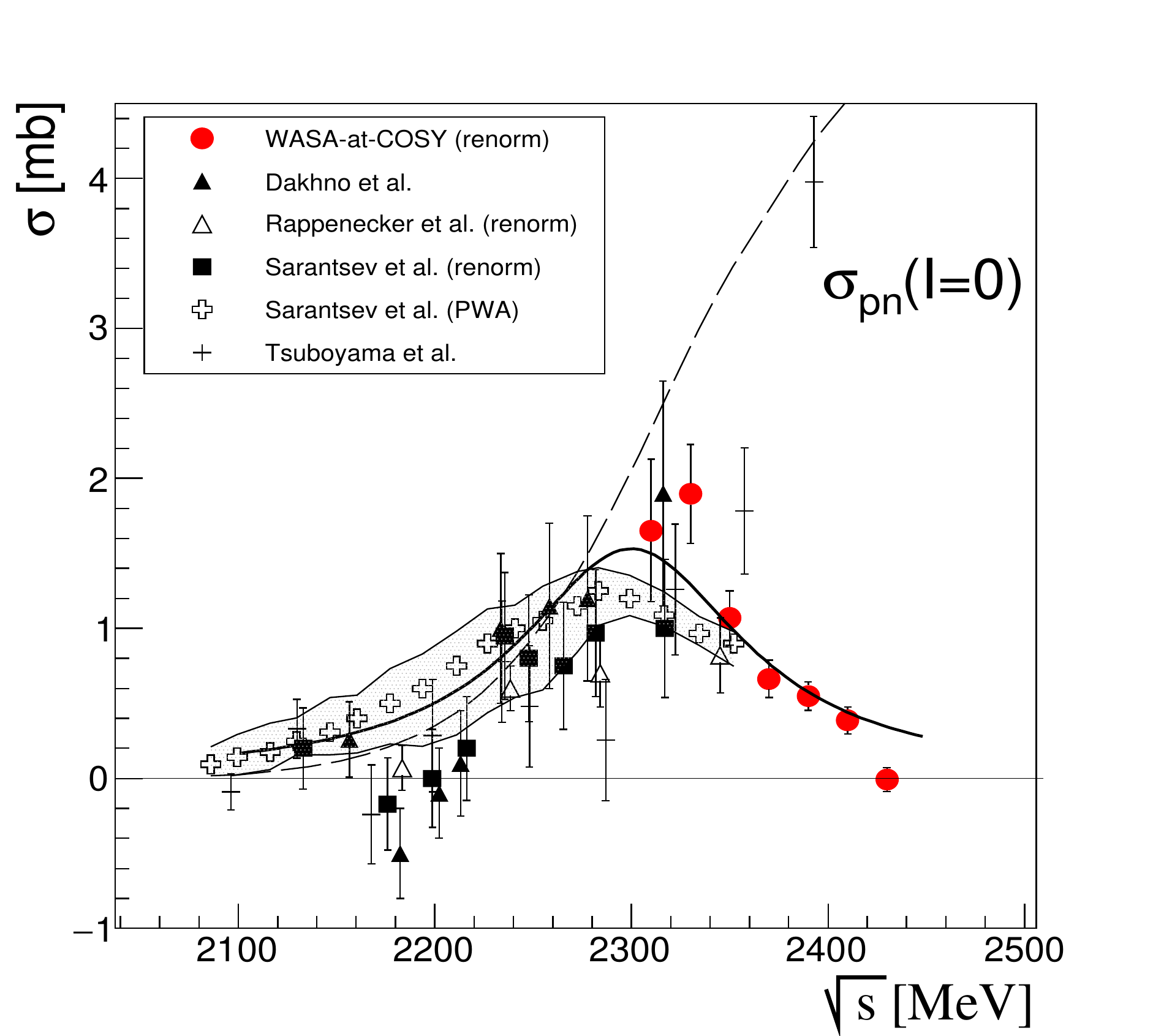}
\caption{\small (Color online) 
The same as Fig.~5, but with renormalized results from
WASA-at-COSY \cite{NNpi,NNpicorr} and Refs.~\cite{Sarantsev,Rappenecker}, see text, and the PWA results of
Ref.~\cite{AS} (open crosses with hatched band).
The solid line represents a Lorentzian fit with $m =$ 2310 MeV and
$\Gamma =$ 150 MeV.
The dashed line shows the expected energy
dependence based on $t$-channel Roper  excitation \cite{NNpi,Luis} adjusted in
height arbitrarily to the data point at $\sqrt s$ = 2260 MeV.
}
\label{fig5}
\end{figure}

In order to visualize, how this bell-shaped isoscalar cross section
evolves, we inspect again Figs.~1 and 3, the energy excitation function of the $pp
\to pp\pi^0$ and $pn \to pp\pi^-$ cross sections, where the isoscalar part
originates from. These cross sections are connected by the isospin relation
given in eq.~(1),
where $\sigma(pp \to pp\pi^0)$ is a purely isovector contribution  of 
$NN$-induced single-pion production. The Valencia model calculations for
$t$-channel $\Delta$ excitation reproduce
this isovector contribution very well for incident energies $T_p >$ 1 GeV ---
both in total (short-dashed line in Fig.~1) and differential cross sections
(Fig.~3 of Ref. \cite{NNpi}). In
the region $T_p = 0.5 - 1.0$ GeV, however, the calculated cross sections
come out much too low. This is understandable, since these calculations do not
include the isovector $\Delta N$ dibaryon excitations with $I(J^P)$ = $1(0^-)$,
$1(2^-)$, $1(2^+)$ and $1(3^-)$ fed by the $^3P_0$, $^3P_2$, $^1D_2$ and $^3F_3$
$pp$-partial waves in an $s$-channel resonance process. Among these $^3P_2$
gives the by far largest contribution to the $pp \to pp\pi^0$ cross section
\cite{AS}. 

Here we are interested just in a simple pragmatic description of
the isovector single-pion production cross section for application in
eq.~(1). Hence we represent these isovector dibaryon excitations conveniently
by a Lorentzian centered at $\sqrt s$ = 2200 MeV with a width of 90 MeV and a
height of about 1 mb (long-dashed curve in Fig.~1), in order to obtain a
reasonable description of the $pp \to pp\pi^0$ cross section
\footnote{According to 
  Ref.~\cite{AS} the $^3P_2$ partial wave provides the by far largest
  contribution to the total cross section with about 1.5 mb. According to 
  Ref.~\cite{Kukulindpi+} not all of the partial wave contribution leads to
  $s$-channel dibaryon formation. Hence a total dibaryon resonance
  contribution of 1 mb appears to be at least qualitatively quite reasonable.
 % Inferring the  resonance 
 % contributions from those of the dibaryon excitations to the $pp \to d\pi^+$
 % cross section, see, {\it e.g.}, Ref.\cite{Kukulindpi+}, the partial 
%suppression of the $I(J^P)$ = $1(2^+)$ and $1(3^-)$ excitations in the
%$pp\pi^0$ channel and the additional contribution of the $I(J^P)$ = $1(0^-)$
%excitation the total resonance contribution of 1 mb at 2200 MeV appears at
%least qualitatively quite reasonable
}. Adding up both contributions gives the dash-dotted curve in Fig.~1, which
provides a very reasonable phenomenological representation of the isovector
single-pion production in the $pp\pi^0$ channel.

Next we consider the $pp\pi^-$ channel, which is isospin mixed. Its isovector
part is given by half of the $pp \to pp\pi^0$ cross section as
illustrated in Fig.~3 by the dash-dotted line. It describes the data in
this channel reasonably up to $\sqrt s \approx$ 2.2 GeV. Beyond this energy
the data exhibit a bell-shaped surplus of cross section, which has to be
purely isoscalar according to eq. (1) and which is well accounted for by the
Lorentzian obtained by the fit to the full isoscalar $pn$-initiated
single-pion production cross section displayed in Fig.~6.

We note in passing that the excursion of the WASA-at-COSY data point at $\sqrt
s$ = 2.32 GeV seen in Fig.~3 and which was focused on in Ref.~\cite{EO} could,
indeed, suggest a tiny narrow structure on 
top of the broad isoscalar Lorentzian. However, the WASA-at-COSY results are
plotted there with statistical uncertainties only. Since this particular data
point is close to the end of the available quasifree regime, systematics such
as model dependence add uncertainties in the range of 5 $\%$ as indicated in
Fig.~2 of Ref.~\cite{NNpi}. Hence this excursion is of no particular
significance. 

In order to learn more details about the nature of the bump structure in the
isoscalar cross section, we 
consider next differential cross sections. In Fig.~6 of
Ref.~\cite{NNpi} the isoscalar $N\pi$-invariant mass spectrum is shown. It
exhibits essentially a single pronounced structure, which peaks at $m \approx$
1370 MeV revealing a width of $\approx$ 150 MeV. This structure emerges well  
above the isovector $\Delta$ excitation (which is filtered out by the isospin
condition) and is located already in the region of the Roper excitation, which
is of both isoscalar and isovector nature. Since in the isoscalar
$N\pi$-invariant mass spectrum the strength is accumulated at the highest
masses available in the reaction process, it follows by kinematics that
the strength in the associated $pp$-invariant mass ($M_{pp}$) spectrum has to
concentrate there at lowest masses. Since such a spectrum was not shown in
Ref.~\cite{NNpi} we plot it now in Fig.~7. The phase-space distribution, which
represents a plain $s$-wave distribution, is indicated by the (yellow) shaded
region. 

\begin{figure} 
\centering
\includegraphics[width=0.89\columnwidth]{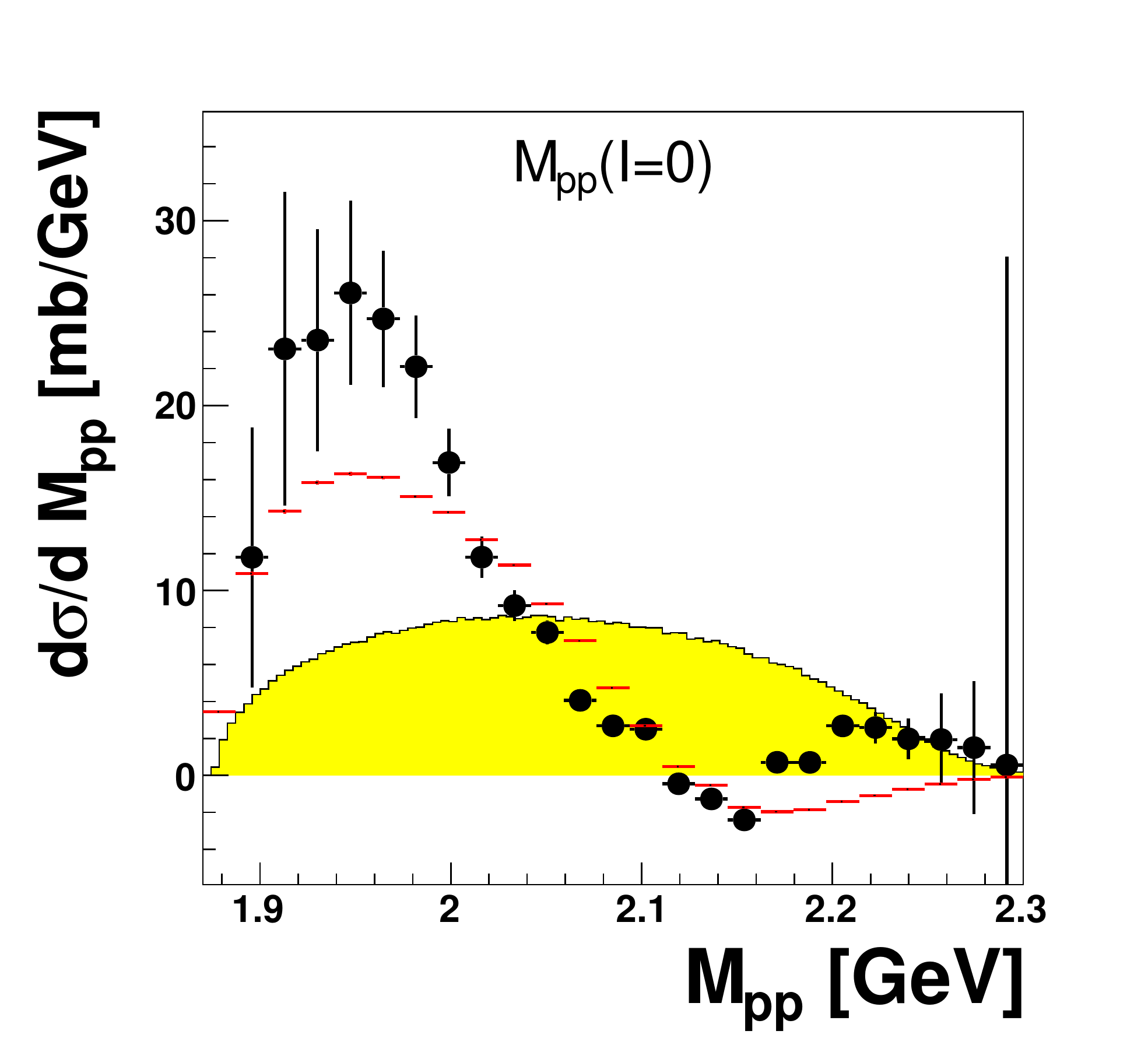}
\caption{\small (Color online)
  Isoscalar $pp$ invariant mass spectrum $M_{pp}(I=0)$ obtained from the
  difference of the corresponding distributions in the $pp \to pp\pi^0$ and
  $pn \to pp\pi^-$ reactions by use of eq.~(1). The phase-space distribution is
  indicated by the (yellow) shaded region. The red dashed histogram gives a
  $t$-channel calculation for Roper excitation. 
}
\label{fig7}
\end{figure}

%We demonstrate this by the corresponding
%Dalitz plot, which we obtain by performing the difference of the Dalitz plots
%for the $pp\pi^0$ and $pp\pi^-$ channels according to eq.~(1).
%From the Dalitz plot for the isoscalar channel
We see that strength
accumulates at lowest $M_{pp}$ values with practically no strength above
$M_{pp} >$ 2050 MeV $= 2m_p + 170$ MeV. From a simple semi-classical estimate
for the centroids of $S$-, $P$- and $D$-wave distributions we get $M_{pp} =
2m_p$, $2m_p + 40$ MeV and $2m_p + 120$ MeV, respectively. {\it I.e.}, the
observed $M_{pp}$ distribution is in accord with $S$- and $P$-waves, but not
with a dominance of $D$-waves as assumed in Ref.~\cite{EO}.  Our finding is,
however, in accord with the result of the partial-wave analyses in
Ref.~\cite{AS}, where $^1S_0$ and $^3P_1$ have been identified as the dominant
$pp$ partial waves in the exit channel. 

$^1S_0$ and $^3P_1$ partial waves incident on the second interaction
process $pp \to d\pi^+$ reduce its cross section by about a
factor of 30 compared to what is assumed in Ref.~\cite{EO}. By using in
addition the proper values for height, peak energy and width of the
isoscalar $np(I=0) \to pp\pi^-$ cross section the sequential single-pion
production ansatz leads to a structure in the $d\pi^+\pi^-$ channel, which is
broader by a factor of two and smaller by about two orders of magnitude than
calculated in Ref.~\cite{EO}, {\it i.e.}, sequential single-pion production is
by no means an explanation for the observed $d^*(2380)$ signal in this channel.

In order to demonstrate the invalidity of the sequential single-pion
production ansatz in this context by yet another example, let us consider now
the isovector part of the $np \to pp\pi^-$ reaction instead of the isoscalar
part. In this case we deal with the two-pion production 
process $np(I=1) \to d\pi^+\pi^-$. Since the isovector part of the 
$pp\pi^-$ channel is larger than its isoscalar 
part by roughly a factor of four near the energy of the $d^*(2380)$ peak (see
Fig.~3), we 
would expect the cross section for the isovector part of the $d \pi^+\pi^-$
channel to be larger than its isoscalar part by just this factor at the
position of $d^*(2380)$. In reality its is smaller by a factor of ten
\cite{isofus} and the sequential single-pion production ansatz fails again
vastly.

\section{The Roper Excitation in $NN$-induced Single-Pion Production and the
  issue of possible $N^*N$ states}

Ever since its first detection by L. D. Roper in 1964 \cite{Roper} the
$N^*(1440)$ resonance has been heavily debated concerning its nature. The
finding that it is in principle of a two-pole nature \cite{Arndt4} increases its
complexity discussed in many subsequent studies
\cite{Cutkosky,Arndt5,Doering,Suzuki}. 

In contrast to the $\Delta$ excitation, the Roper excitation $N^*(1440)$ in
general does not produce very eye-catching structures in hadronic
reactions. Usually it appears quite 
hidden in the observables and in most cases can be extracted from the data
only by sophisticated analysis tools like partial-wave decomposition. As an
exception appears here the $pn$-induced isoscalar single-pion production,
where it can be observed free of the usually overwhelming isovector $\Delta$
excitation as demonstrated by 
recent WASA-at-COSY results for the $pn \to (NN\pi)_{I=0}$ reaction
\cite{NNpi}. The primary aim of this experiment was the search for a decay
$d^*(2380) \to (NN\pi)_{I=0}$. But since the measured energy range covers also
the region of the Roper excitation, it fits the purpose of the topic of this
work, too.

The extracted values for mass and width of the structure observed in the
isoscalar nucleon-pion invariant-mass spectrum (Fig.~6 of Ref. \cite{NNpi})
appear to be  compatible with the pole values for the Roper
resonance deduced in diverse $\pi N$ and $\gamma N$ studies \cite{PDG}. Our
values for the Roper peak are in reasonable agreement, too, with earlier 
findings from hadronic $J/\psi \to \bar{N} N\pi$ decay \cite{BES} and $\alpha
N$ scattering \cite{Morsch1,Morsch2}. However, our values deviate
substantially from its Breit-Wigner 
values, which for the Roper resonance are quite different and which should be
the standard to compare with. With regard to its Breit-Wigner mass the Roper
resonance appears to be bound by about 70 MeV within the $N^*N$ system. Such a
binding then also explains naturally its observed reduced width of 150 MeV, 
since the Roper width is strongly momentum dependent due to its $N\pi$
$p$-wave nature. 

Since at threshold the conventional $t$-channel Roper excitation can be
expected to be produced in $S$-wave relative to the 
other nucleon and since the pion from the Roper decay is emitted in relative
$p$-wave, we would conventionally expect a threshold behavior for the energy
dependence of $pn \to (NN\pi)_{I=0}$  cross section like that
for pion $p$-waves as born out by the calculations for $t$-channel Roper
excitation in the framework of the modified Valencia model \cite{NNpi,Luis}
--- in Figs.~5, 6 arbitrarily adjusted in height to the data point at $\sqrt s
\approx$ 
2260 MeV and displayed by the dashed line. The data presented there 
follow this expectation by exhibiting an 
increasing cross section with increasing energy up to about $\sqrt s \approx$
2.30 GeV. Beyond that, however, the data fall in cross section in sharp
contrast to the expectation for a $t$-channel production process. The
observed behavior rather is in agreement with a $s$-channel resonance process as
expected for the formation of a dibaryonic state near the $N^*N$ threshold.

If we combine this dibaryon hypothesis with the result of the partial-wave
analysis \cite{AS} for the isoscalar single-pion production, then the observed
bump structure must consist actually of two resonances: one resonance, where
$N$ and $N^*$ are in relative $S$-wave yielding $I(J^P) = 0(1^+)$ and
connected to the coupled $^3S_1 - ^3D_1$  $np$ partial waves --- and
one resonance, where $N$ and $N^*$ are in relative $P$-wave yielding $I(J^P) =
0(1^-)$ and connected to the $^1P_1$ partial wave. On a first glance it might
not appear very convincing that two resonances sit practically on top of each
other and produce thus just a single resonance-like structure in the total cross
section. But exactly such a scenario is observed also near the $\Delta N$
threshold, where the isovector $0^-, 2^+, 2^-$ and $3^-$ states happen to have 
similar masses with mass differences small compared to their width. And since
the  width of the $N^*N$ states is still substantially larger than that of
the $\Delta N$ states, small mass differences are washed out in the summed
shape. We note that $1^+$ and $1^-$ constitute the only possible $J^P$
combinations for isoscalar $S$ and $P$ waves.

In the following we examine, whether this dibaryon hypothesis leads to any
conflicts with regard to unitarity, decay properties and poles in elastic $np$
scattering.

\subsection{Relation to Isoscalar Two-Pion Production}
 
Since the Roper resonance decays in addition via two-pion emission, the same
should be valid also for the $N^*N$ configuration. Indeed, there is an
indication of such a decay in 
the $pn \to d\pi^0\pi^0$ reaction, which might solve another puzzling 
problem. Whereas the data for this reaction can be reasonably well described
by a simple Breit-Wigner ansatz with momentum-independent widths, the
description worsens on the low-energy side, if we apply a sophisticated
momentum-dependent ansatz for the widths \cite{abc}.

The situation is shown in Fig.~8, where the energy dependence of the
total cross section for the $pn \to d\pi^0\pi^0$ reaction is plotted. Since the
conventional background of $t$-channel processes is particularly low in this
reaction channel, it is so-to-speak the "golden" channel for the observation of
the $d^*(2380)$ dibaryon resonance. The solid line represents the calculated
$d^*$ excitation taking into account the momentum dependence of its width in
very detail \cite{abc}. This theoretical curve describes the data very well
except in the low-energy tail of $d^*(2380)$ around $\sqrt s$ = 2.3 GeV, where
it clearly  
underpredicts the data. If we plot the difference between data and
calculation by the (black) filled dots in Fig.~8, then we note a bell-shaped
distribution, the right-hand side of which being strongly dependent on the
details (mass, width) of the $d^*(2380)$ resonance curve. Associating this
distribution with a contribution from the possible $N^*N$ structure we can
deduce a peak cross
section of roughly 25 $\mu$b at 2.3 GeV for its two-pion decay into the
$d\pi^0\pi^0$ channel. Consequently we expect a contribution of such a
two-pion decay of the $N^*N$ system also in the other
two-pion production channels with isoscalar contribution, which are the channels
$d\pi^+\pi^-$, $pn\pi^+\pi^-$ and $pn\pi^0\pi^0$. By isospin relations the
isoscalar Roper contribution in the first two reactions is twice that in each
of the channels $d\pi^0\pi^0$ and $pn\pi^0\pi^0$. Here we also assume that the
branching into $d\pi^+\pi^-$ ($d\pi^0\pi^0$) and $pn\pi^+\pi^-$
($pn\pi^0\pi^0$) channels is essentially identical --- as is the case for
$d^*(2380)$ \cite{BR,FW,AO}. Altogether these contributions add then up to a
total of roughly 150 $\mu$b.

In the two-pion decay of the $N^*N$ systems with $J^P = 1^+$ the emitted
particles are in relative $S$-wave to each other. In case of $J^P = 1^-$ the
pions are in $P$-wave relative to the deuteron. Both contributions appear
summed up in the angular distributions for deuterons and pions. Hence we
expect only mildly curved angular distributions, which
actually agrees with the observations for $\sqrt s <$ 2.34 GeV, where the
$d^*(2380)$ contribution is still small \cite{Internal}.

\subsection{Branching Ratios of putative $N^*N$ Resonances}

Having identified all inelastic decay channels we can extract now the
branching ratios for $(N^*N)_{I(J^P)=0(1^+)} \to NN, NN\pi$ and $ NN\pi\pi$ in
analogy to what was done for $d^*(2380)$ \cite{BR}.

For a $J = 1$ resonance formed in $pn$ collisions at 2315 MeV the unitarity
limit is given by \cite{BR} 

\begin{eqnarray}
  \sigma_0 = \frac{4\pi}{k_i^2}\frac{2J+1}{(2s_p+1)(2s_n+1)} = 8~mb,
\end{eqnarray}

where $k_i$, $s_p$ and $s_n$ denote the initial center-of-mass momentum, the
proton and the neutron spin, respectively. The branching ratio for the decay
into the elastic channel, $BR_i = \Gamma_i/\Gamma$ with $\Gamma_i$ and
$\Gamma$ denoting the decay widths into the initial channel and the total
width, respectively, is then given by  \cite{BR} 

\begin{eqnarray}
  BR_i = \frac{1}{2} - \sqrt{\frac{1}{4} - \frac{\sigma_{pn \to
  N^*N}(peak)}{\sigma_0}}.
\end{eqnarray}

From the partial-wave analysis of Ref.~\cite{AS} we infer  that about 25$\%$
(75$\%$) of $\sigma_{pn \to NN\pi}(I=0)$ contributes to the $1^+$ ($1^-$)
state with a peak cross section of 0.3 (1.0) mb. This leads then in eq.~(3) to
$BR_i(1^+) = 0.04(2)$ and $BR_i(1^-) = 0.15(3)$, respectively. The branchings
into $NN\pi$ and $NN\pi\pi$ channels are then 85(10)$\%$ and 11(2)$\%$,
respectively, for the $1^+$ state. For the $1^-$ state these numbers get
75(15)$\%$ and  10(2)$\%$, respectively. The estimated uncertainties quoted in
brackets include those from the partial-wave analysis and a 20$\%$ uncertainty
in the absolute scale of the isoscalar cross section.

%where the uncertainty results largely from the 30$\%$ uncertainty in the absolute
%scale of the deduced isoscalar cross section \cite{NNpi}. The branching into the
%$NN\pi$ channel is then about 75(20)$\%$ and that into the $NN\pi\pi$ channel
%about 8(3)$\%$.

%For this estimate we have to consider that there should be also a nonresonant
%background from conventional $t$-channel pion production. From inspection of
%Fig.~1 we may infer that this contribution cannot be larger than 0.3 mb in
%$\sigma_{pn}(I=0)$.  
% This number corresponds to a 0.1 mb contribution in the $pp\pi^-$ channel.
%Since the $t$-channel Roper excitation contributes to isoscalar and
%$isovector pion production with equal weights, we may inspect the  $pp \to
%pp\pi^0$ reaction for a rough cross check of this number. From the analysis of
%its differential cross sections \cite{NNpi}, in particular 
%its invariant-mass distributions, we can estimate an upper limit of 0.4 mb
%for the contribution of the total (nonresonant and resonant, see next section)
%isovector  Roper excitation --- a number, which is in the same order of
%magnitude and which means that, indeed, the $t$-channel Roper excitation cross
%section must be small. 

Recently Kukulin {\it et al.} \cite{NstarN} have predicted a $I(J^P) = 0(1^+)$
resonance based on the analysis of the $^3S_1$ $NN$-partial wave within the
dibaryon-based $NN$-interaction model \cite{KukulinNN,Kukulin1}, where also a
short preview of this work was provided. The Argand plot of the calculated
$^3S_1$ partial wave (Fig.~6 in Ref.~\cite{NstarN}) shows a resonance
circle with diameter of about 0.09, which according to H\"ohler \cite{Hoehler}
corresponds just to the elastic branching ratio $BR_i$. Though this value means
already a very small elasticity, it is still somewhat larger than we obtain
here. 

\begin{figure} 
\centering
\includegraphics[width=0.99\columnwidth]{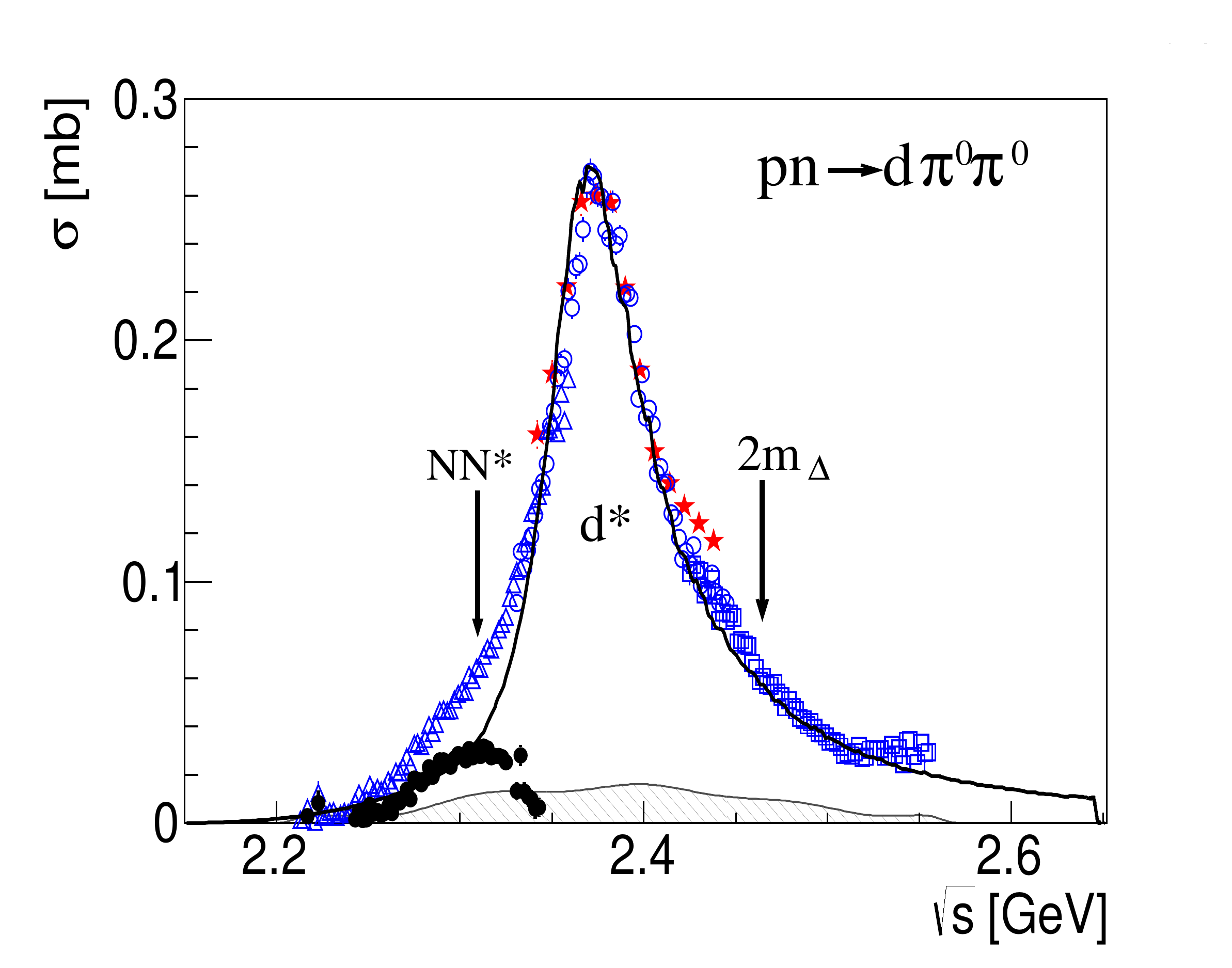}
\caption{\small (Color online) 
Energy dependence of the total cross section for
  the $pn \to d\pi^o\pi^0$ reaction as measured by WASA-at-COSY. The blue open
  symbols represent the data of Ref.~\cite{prl2011} normalized to the data
  (red stars) of Ref.~\cite{isofus}. The hatched area gives an estimate
  of systematic uncertainties. The solid curve displays a calculation of the
  $d^*$ resonance with momentum-dependent widths \cite{abc}. It includes both
  Roper and $\Delta\Delta$ $t$-channel excitations as background
  reactions. The black filled dots show the difference between data and this
  calculation in the low-energy tail of $d^*(2380)$.
}
\label{fig8}
\end{figure}

\subsection{Poles of $N^*N$ resonances in elastic $np$ scattering}

In principle, the poles of such $N^*N$ resonances  should be sensed in a 
partial wave analyses of elastic $np$ scattering. At a first glance, the
situation appears to be similar to that for the meanwhile established dibaryon
resonance $d^*(2380)$, where only the measurement of the analyzing power of $pn$
scattering in the region of this resonance could reveal its pole in the $^3D_3
- ^3G_3$ coupled partial  waves \cite{np,npfull,pnxsection}. Due to their large
angular momenta these 
partial waves have a large impact on the analyzing power. And since the
analyzing power consists of just 
interference terms, this observable is very suitable to reveal substantial
effects even from small resonance admixtures in partial waves. In case of the
$I(J^P) = 0(1^+)$ resonance candidate 
we deal here with a $S$-wave resonance, which makes no contribution to the
analyzing power. Hence this key observable for revealing small contributions
from resonances is not working here. In addition, the large total width of
these $N^*N$ resonances combined with small a elasticity $BR_i$ increase the
difficulty to reveal their poles by elastic scattering. Even the
dedicated dibaryon search by high-resolution energy scans of $pp$ elastic
scattering with the EDDA detector at COSY was restricted to the search of
narrow resonances only \cite{EDDA}. 

Though it seems that we have no suitable handle to reveal the pole of such a
$S$-wave resonance by partial-wave analyses of elastic scattering data, their
imprint on the $^3S_1$ partial wave due to intermediate dibaryon
formation in the $s$-channel $NN$-interaction has been shown to be
significant. In Ref. \cite{NstarN} it has been demonstrated that this
resonance leads to a quantitative reproduction of the empirical values for
coupled $^3S_1$-$^3D_1$ partial waves up to 1 GeV obtained in SAID partial-wave
analyses \cite{SAIDPW}. Conversely, the successful
description of these partial waves also means that our finding about this
resonance is not in conflict with elastic scattering data.   

For the $I(J^P)=0(1^-)$ state the situation appears perhaps a bit more
promising. It 
is true that again the large width of this state hampers any detection of its
resonance signal in elastic $pn$ scattering enormously, but its increased
elastic branching of about 15$\%$ is in favor of a better sensible signal
there. Unfortunately the SAID single-energy solutions stop at 1.1 GeV and
hence cover only the low-energy tail of this resonance candidate.

%In both cases, in the $I(J^P)=0(1^+)$ and in the $I(J^P)=1(0^+)$ case, the
%Roper excitation appears with values for mass and width compatible with its
%deduced pole parameters, but not with its Breit-Wigner values of $m \approx$
%1440 MeV and $\Gamma \approx$ 350 MeV \cite{PDG}, which are observed in $\pi
%N$ and $\gamma N$ scattering experiments. As noted above already, our
%observation coincides with that in $\alpha N$ scattering. A possible
%explanation might be given by the formation of $N^*N$ resonances in
%nucleon-induced reactions, where due to a binding of about 70 MeV the effective
%Roper resonance mass is reduced to 1370 MeV --- thereby also reducing its width
%due to its strong $p$-wave momentum dependence to about 150 MeV. This is also
%in accord with our observation that the decay branching into the two-pion
%channel is smaller than that of the bare Roper decay \cite{PDG}.

\section{Conclusions}

We have reanalyzed the situation of $NN$-induced isoscalar single-pion
production. The total cross section data exhibit a bump-like energy
dependence, which can be described by a Lorentzian with mass 2310 MeV and
width of about 150 MeV. This is at variance with a width as narrow as 70 MeV
assumed in Ref.~\cite{EO}. Also, the emitted $pp$ pair is dominantly in
relative $S$ and $P$ wave, but not in the $^1D_2$ wave, as assumed in the
sequential single-pion production ansatz of Ref.~\cite{EO}. In consequence the
resonance signal calculated by this ansatz for the $pn \to d\pi^+\pi^-$
reaction gets broader by a factor of two and smaller by about two orders of
magnitude than calculated in Ref.~\cite{EO}.

The fact that the observed isoscalar $M_{N\pi}$ spectrum accumulates most of
its strength in the region of the Roper resonance suggests that the observed
bump is of $N^*N$ nature. Taking into account the results of the partial-wave
analysis of Ref.~\cite{AS}, then this resonance-like structure contains
actually two isoscalar resonances, one 
with $J^P = 1^+$ and the other one with $J^P = 1^-$. This are also
the only two possibilities to form resonances in isoscalar $S$- and $P$-wave $NN$
scattering. The situation appears similar to the one observed near the
$\Delta N$ threshold, where several resonances have been found, which all have
similar mass and width.

From the energy dependence of $NN$-induced isoscalar single-pion and isovector
double-pion production we see that both isospin-spin combinations in the
$N^*(1440)N$ system lead possibly to dibaryonic states in the Roper
excitation region --- analogous to the situation at the $\Delta$
threshold. However, compared to the situation there the Roper excitation cross
sections discussed here are small. Also, since these structures decay
mainly into inelastic channels, their poles are hard to be sensed in
partial-wave analyses of elastic scattering. Nevertheless, their effect on the
$^3S_1$ and $^3P_1$ $NN$ partial waves have been shown to be important 
in the $NN$-interaction model of Kukulin {\it et. al.}
\cite{KukulinNN,Kukulin1}, where the short-range part of the $NN$-interaction
is represented by $s$-channel dibaryon formation in the various low-$L$
partial waves based on ideas given in Ref. \cite{KukulinAnn}.

\section{Acknowledgments}

We are indebted to V. Kukulin\footnote{deceased} and M. Platonova for valuable
discussions and to L. Alvarez-Ruso for using his code.  We acknowledge
valuable discussions with E. Oset, A. Gal and I. Strakovsky.
This work has been supported by DFG (CL 214/3-3).

\end{document}